\documentclass[mathleft
]{an}
\usepackage{graphicx}
\usepackage{times}
\usepackage{amsmath}
\begin{document}

\Pagespan{1032}{1035}
\Yearpublication{2012}%
\Yearsubmission{2012}%
\Month{December}%
\Volume{333}%
\Issue{10}%
\DOI{10.1002/asna.201211821}

\title{Asteroseismology and Magnetic Cycles}

\author{A.R.G. Santos\inst{1,2}\fnmsep\thanks{\email{asantos@astro.up.pt}\newline},
M. S. Cunha \inst{1} \and J.J.G. Lima \inst{1,2}}
\titlerunning{Asteroseismology and Magnetic Cycles}
\authorrunning{A.R.G. Santos, M.S. Cunha \& J.J.G. Lima}
\institute{
Centro de Astrof\'{i}sica da Universidade do Porto, Rua das Estrelas, 4150 Porto, Portugal
\and 
Departamento de F\'{i}sica e Astronomia, Faculdade de Ci\^{e}ncias, Universidade do Porto, Rua do Campo Alegre, 687, 4169-007 Porto, Portugal}

\received{2012}
\accepted{2012}
\publonline{2012 Dec 3}


\abstract{%
Small cyclic variations in the frequencies of acoustic modes are expected to be a common phenomenon in solar-like pulsators, as a result of stellar magnetic activity cycles.
The frequency variations observed  throughout the solar and stellar cycles contain information about structural changes that take place inside the stars as well as about variations in magnetic field structure and intensity. The task of inferring and disentangling that information is, however, not a trivial one. 
In the sun and solar-like pulsators, the direct effect of the magnetic field on the oscillations might be significantly important in regions of strong magnetic field (such as solar- / stellar-spots),  where the Lorentz force can be comparable to the gas-pressure gradient. Our aim is to determine the sun- / stellar-spots effect on the oscillation frequencies and attempt to understand if this effect contributes strongly to the frequency changes observed along the magnetic cycle. The total contribution of the spots to the frequency shifts results from a combination of direct and indirect effects of the magnetic field on the oscillations. In this first work we considered only the indirect effect associated with changes in the stratification within the starspot. Based on the solution of the wave equation and the variational principle we estimated the impact of these stratification changes on the oscillation frequencies of global modes in the sun and found that the induced frequency shifts are about two orders of magnitude smaller than the frequency shifts observed over the solar cycle.}

\maketitle

\section{Introduction}

The level of magnetic activity in stars varies periodically with time. The best example of this is provided by the sun and the most obvious sign of the solar activity is the periodic variation of the number of sunspots over the so-called solar cycle. The magnetic activity in the sun and other stars also influences the properties of the global waves that propagate within them. In the sun, the variations of the oscillation properties along the solar cycle are well documented and studied (e.g. Libbrecht \& Woodard 1990; Chaplin et al. 2007): during the solar maximum the frequencies of acoustic modes increase and the amplitudes decrease. Frequency shifts were also discovered in a CoRoT star (Garcia et al. 2010) and are expected to be found in stars observed by Kepler.

Sunspots induce direct and indirect effects on the oscillation properties. The total contribution combines these two components. The direct effect results from the action of the Lorentz force on the perturbations and the indirect effect is due to the magnetically induced changes in the stratification within the spot and due to the effect of the magnetic field on the convective flows, which affects the efficiency of convection and the turbulent velocities.

In this paper we describe a possible approach to calculate some of these effects and present the results for the indirect contribution associated with changes to the star's stratification within starspots. In the following section (Section \ref{sec:hydrodynamics}), we present the wave equation and the method adopted to solve it. To solve the wave equation we need a starspot model. The model used in this work is presented in Section \ref{sec:model}. Finally, we show our results for the case of the sun in Section \ref{sec:results}.

\section{Wave Equation}
\label{sec:hydrodynamics}

In this work we consider only the indirect effect of the magnetic field on the oscillations, in particular that associated with the stratification changes within the starspot, as compared with the spherically symmetric case.

Applying the linear perturbation theory and neglecting the perturbation to the gravitational potential and the direct effect of the magnetic field on the oscillations (i.e., neglecting the terms coming from the Lorentz force), we can derive, in the adiabatic case, the following wave equation
\begin{equation}
\rho_o\dfrac{\partial^2\vec{\xi}}{\partial t^2}=\nabla\left(\gamma p_o\nabla\cdot\vec{\xi}\right)+\nabla\left(\vec{\xi}\cdot\nabla p_o\right)+\nabla\Phi_o\nabla\cdot\left(\rho_o\vec{\xi}\right)\label{eq:waveeq}\end{equation}
\normalsize where $\vec{\xi}$ is the displacement around the equilibrium position, $t$ the time, $\gamma$ the first adiabatic exponent, $\rho$ the density, $p$ the pressure, $\Phi$ the gravitational potential and the index \textit{'o'} denotes the equilibrium quantities. The indirect effect of the magnetic field is seen in this equation in the terms involving $\rho_o$ and $p_o$. These quantities differ from the form they take in the spherically symmetric case, depending on depth and on the distance from the spot's axis according to the model presented in Section \ref{sec:model}.

To solve Equation (\ref{eq:waveeq}) we used cylindrical coordinates ($r$, $\phi$, $z$) and took advantage of the axial symmetry of the spot model (Section \ref{sec:model}). In practice we followed the procedure described below:

\begin{enumerate}
\item We started by neglecting the horizontal derivatives of the displacement by assuming that the displacement varies faster in the vertical direction. We also neglected the horizontal derivatives of pressure and density, solving the equation locally at each distance from the spot's axis ($r$). In this way our problem was reduced to a second order total differential equation for the vertical displacement. This equation was solved in the region of the spot, down to a depth of - 40 Mm (thus, beyond the spot's base), by applying the boundary condition that the Lagrangian pressure perturbation vanishes at the outermost point of our model, which implies $(\nabla\cdot\vec{\xi})=0$, and a normalization condition at the surface.

\item Since the pressure and the density stratifications within the spot depend on $r$, the solution found in point 1. is also dependent on $r$. Thus, we took the approximate solution derived in 1. and computed the horizontal derivatives of the displacement. With this at hand we considered again Equation (\ref{eq:waveeq}), now taking into account also the horizontal derivatives of pressure and density, and derived a new solution for the vertical component of the displacement, as well as the solution for the horizontal component.

\item Finally, with the vertical and horizontal components at hand from the previous iteration we calculated their horizontal derivatives and fed them back into the original equation for a new solution. This last procedure was repeated until the horizontal and vertical components of the displacement remained unchanged.
\end{enumerate}

In practice, the final solution for the vertical component of the displacement differed only very little from the first solution, thus confirming that the terms involving the horizontal derivatives have a relatively small impact on the solutions for the vertical displacement component.

\section{Starspot Model}
\label{sec:model}

In order to determine the contribution to the frequency shifts of the regions of intense magnetic field, we need a starspot model in magnetohydrostatic equilibrium. We use the sunspot model of Khomenko \& Collados (2008) that is constructed in cylindrical coordinates, with $\vec{B}$, $p_o$ and $\rho_o$ dependent on the distance from the spot's axis, $r$,  and on the height measured from the field-free photosphere, $z$. This model combines two other models: the approach of Pizzo (1986) for the photosphere and the model of Low (1980) that is more appropriate in deep layers.

On the outer boundary the combined model merges into a quiet sun nonmagnetic model (Model S of Christensen-Dalsgaard et al. (1996)) matched to the VAL-C model of the solar chromosphere (Vernazza et al. 1981). Since we want to integrate the wave equation to layers that are deeper than the spot's base, we join the pressure and density profiles of the spot into those of the Model S at the lower boundary.
\begin{figure}[h!]
\includegraphics[width=8.2cm]{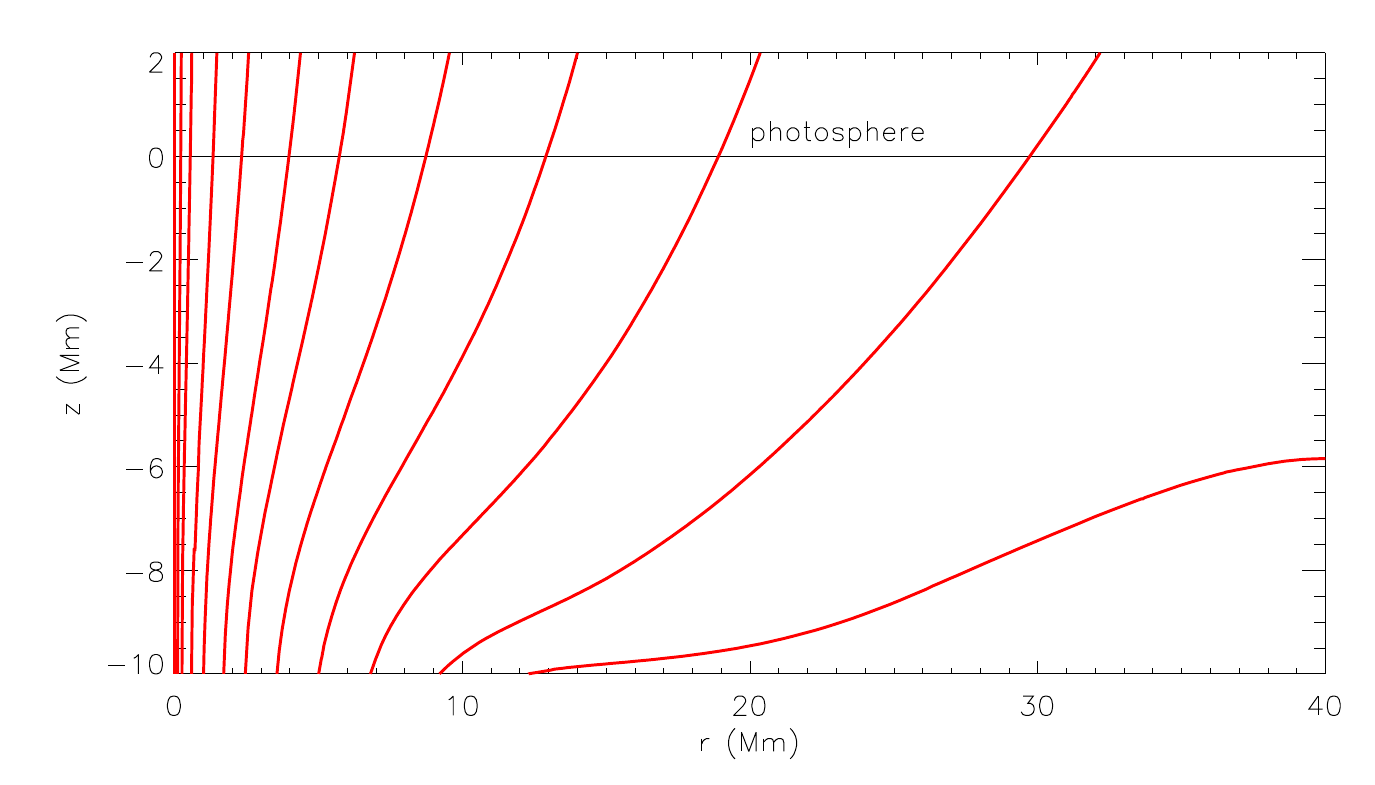}
\caption{Magnetic field lines of Khomenko \& Collados model.}
\label{fig:maglines}
\end{figure}

The magnetic field topology of the sunspot model is shown in Figure \ref{fig:maglines}, where $z=-10$ Mm is the sunspot base and $z=0$ is the photosphere ($z$ and $r$ are the cylindrical coordinates).
\begin{figure}
\includegraphics[width=8.2cm]{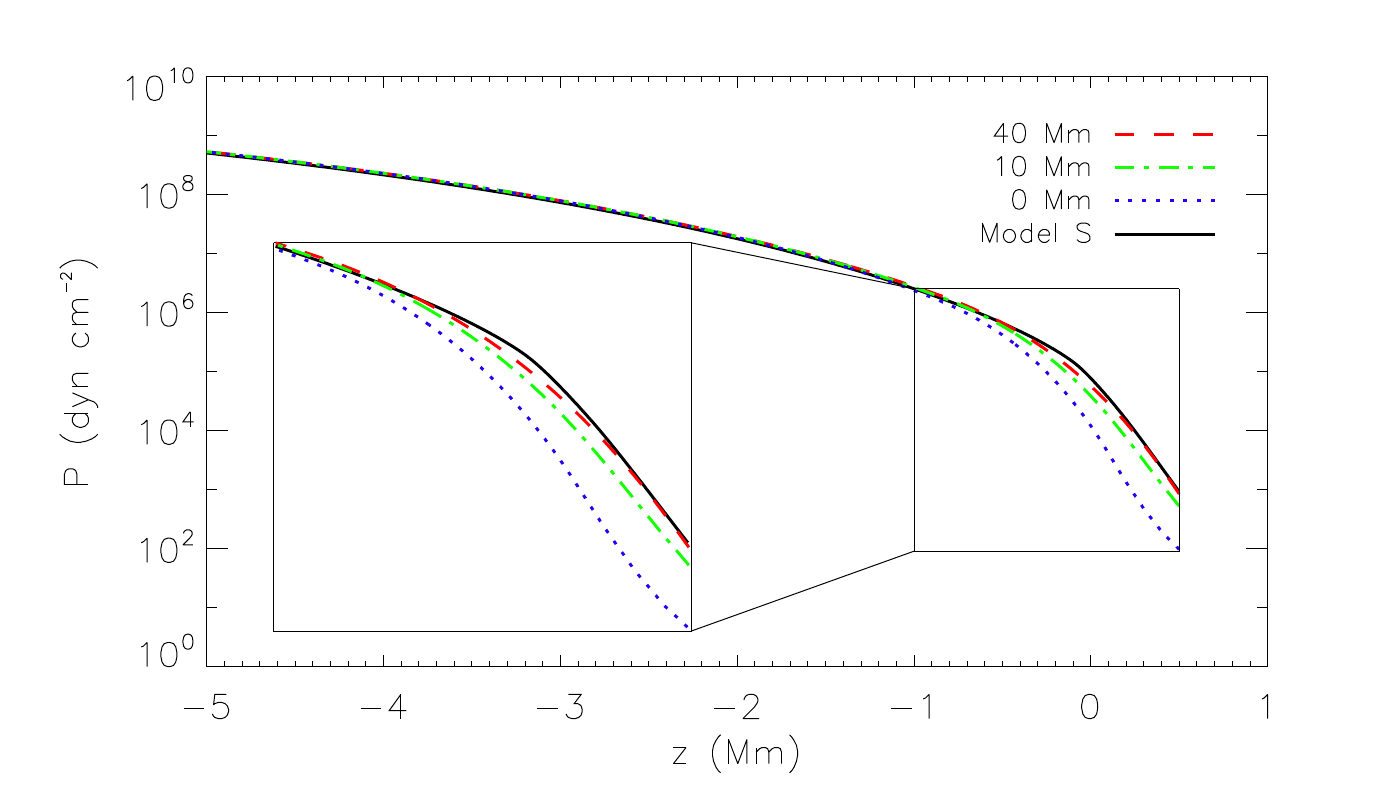}
\caption{Pressure profile of Model S and of Khomenko \& Collados model for three values of the distance $r$ from the sunspot's axis: at the spot's outer boundary (40 Mm), umbra (10 Mm) and at the spot's axis (0 Mm). The larger square shows a zoom of the region inside the smaller square.}
\label{fig:pressure}
\end{figure}

\begin{figure}
\includegraphics[width=8.2cm]{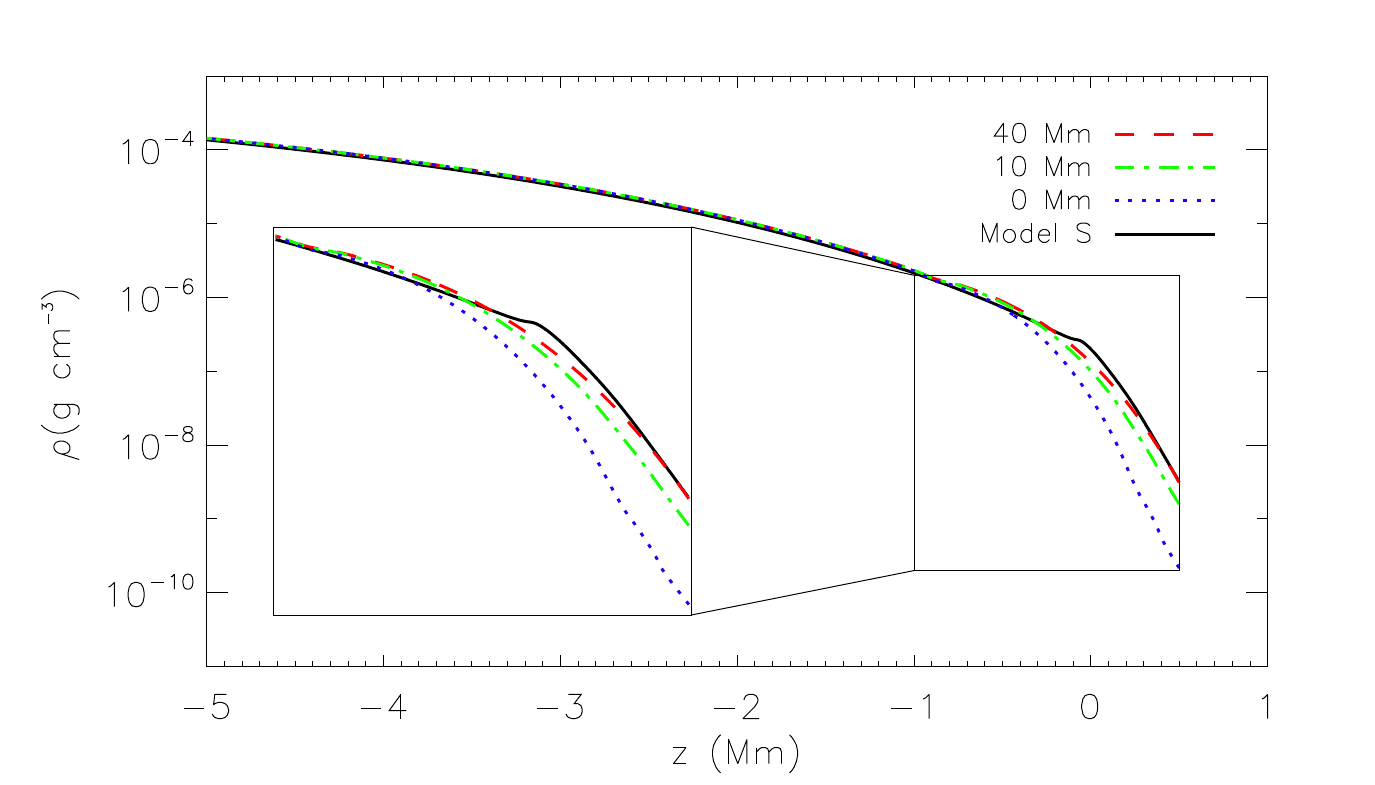}
\caption{Density profile of Model S and of Khomenko \& Collados model for three values of the distance $r$ from the sunspot's axis: at the spot's outer boundary (40 Mm), umbra (10 Mm) and at the spot's axis (0 Mm). The larger square shows a zoom of the region inside the smaller square.}
\label{fig:density}
\end{figure}

We can see through the pressure and density profiles (Figures \ref{fig:pressure} and \ref{fig:density}) that the Khomenko \& Collados model does not merge perfectly to the Model S at large distances from the spot's axis. Since we are interested in investigating the impact of the sunspot structure on the oscillations we will thus take as the reference magnetic-free model the model of Khomenko and Collados at the spot's outer boundary ($r=40$ Mm). Considering Model S as the reference would introduce an artificial difference between the spot and non-spot solutions, which would be associated with the imperfect matching between the stratification at the spot's edge and the stratification of the quiet sun. 

\section{Results of the Indirect Effect of Sunspots}
\label{sec:results}

\begin{figure}[h!]
\includegraphics[width=8.2cm]{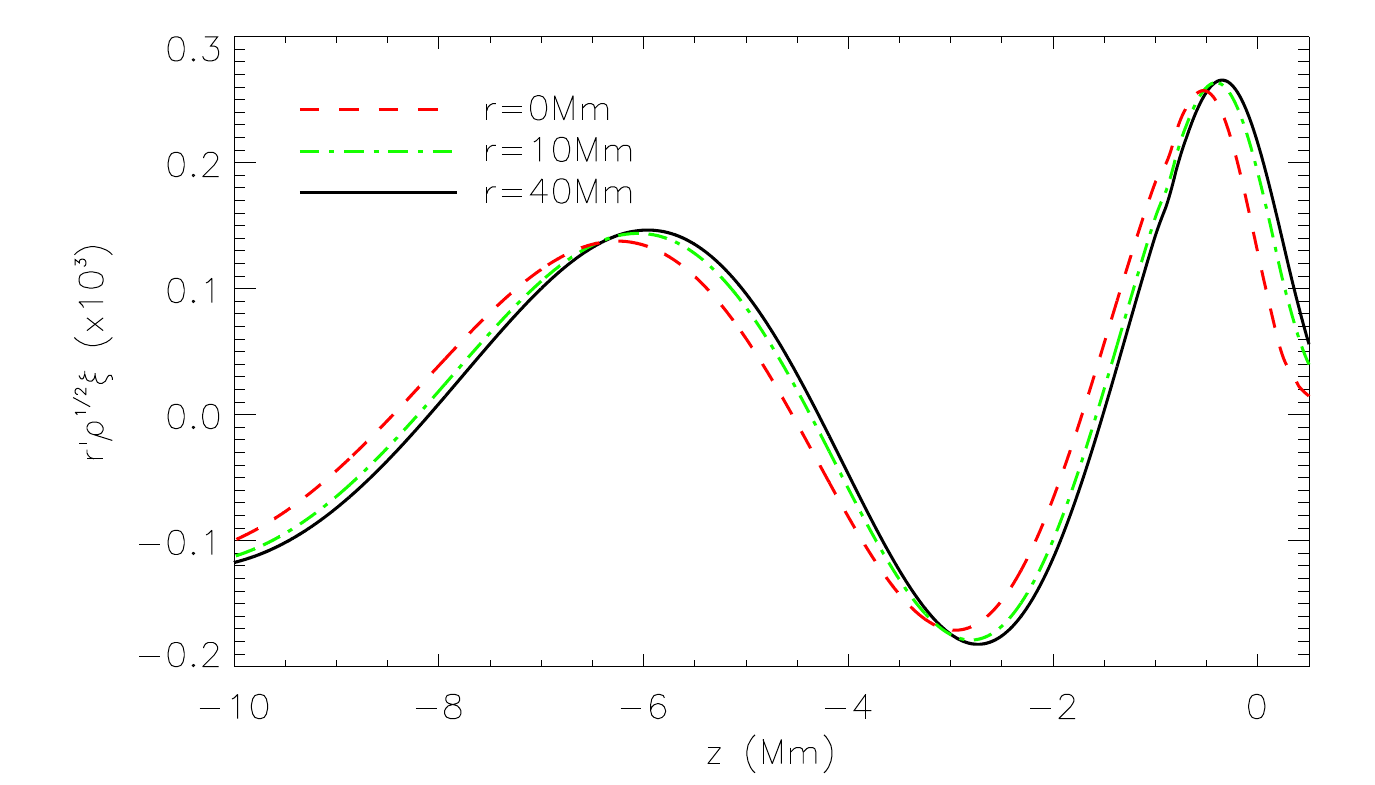}
\caption{Scaled displacement solution ($\nu=3711\hspace{0.1cm}\mu Hz$, $l=0$) for the non-magnetic case at the spot's outer boundary (40 Mm), umbra (10 Mm) and at the spot's axis (0 Mm).}
\label{fig:solutions}
\end{figure}

In order to evaluate the impact on the oscillations of the structural variation within the spot, we computed the solutions of Equation (\ref{eq:waveeq}) for a mode of $l=0$ and several frequencies. In Figure \ref{fig:solutions} we show the solutions for the radial displacement, found following the procedure described in Section \ref{sec:hydrodynamics}, at different distances from the sunspot's axis for $\nu=3711\hspace{0.1cm}\mu Hz$. We emphasize that the differences in these solutions result only from the differences in the stratification. No direct effect of the magnetic field on the oscillations is considered here. 

\begin{figure}
\includegraphics[width=8.2cm]{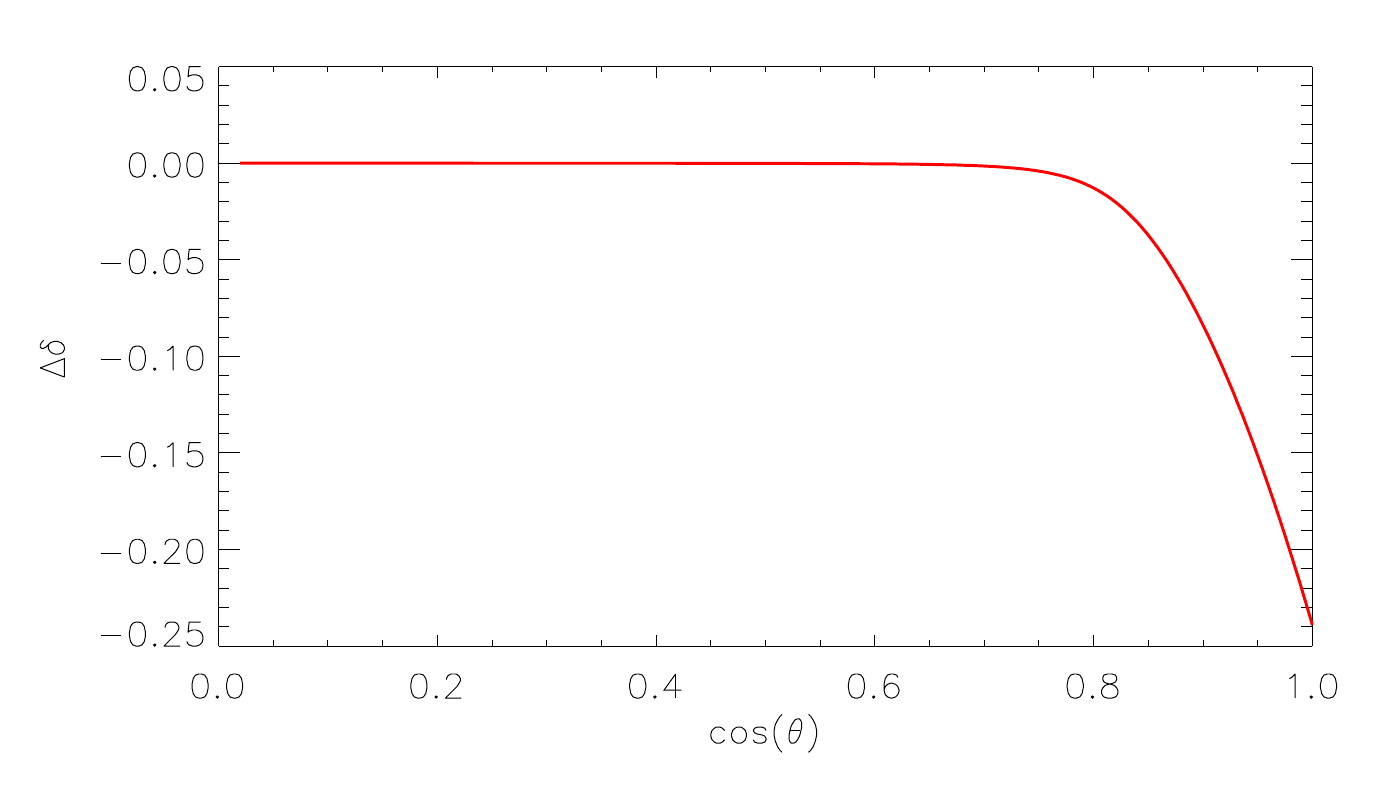}
\caption{Phase difference between the reference solution (at 40 Mm) and the solutions at different distances to 
the spot's axis as a function of the cosine of the photospheric magnetic field inclination angle.}
\label{fig:phasediff}
\end{figure}

With the vertical component of the solution at hand, we compute the phase by matching it to the asymptotic solution (Gough 1993) at each $r$
\begin{equation}
\xi_z\sim\cos\left(\int_z^{z^*}\kappa dz+\delta(z^*,r)\right),
\label{eq:xizfit}\end{equation}
where $z^*$ is the depth at the spot's base, $\delta$ the phase and $\kappa$ is the acoustic wavenumber
\begin{equation}
\kappa^2=\frac{\omega^2-\omega^2_c}{c^2},
\label{eq:kacous}\end{equation}
where $\omega$ is the angular frequency, $\omega_c$ is the critical acoustic frequency and $c$ is the adiabatic sound speed. The phase difference between the reference solution at 40 Mm and the solutions for different distances from the spot's axis is illustrated in Figure \ref{fig:phasediff}, as a function of the cosine of $\theta$, where $\theta$ is the inclination angle of the magnetic field at the photosphere relatively to the vertical axis. As expected, the phase difference changes smoothly from zero, as we approach the spot's axis, becoming significant only in the innermost regions where the pressure and density stratifications are significantly different from their quiet sun counterparts.

\subsection{Induced Frequency Shifts}
\label{sec:freqshifts}

The phase difference computed in Section \ref{sec:results} can be used to compute the frequency shifts of global modes by following the approach of Cunha and Gough (2000). Using a variational principle, these authors have shown that the angular frequency shift is related to the phase difference by the expression
\begin{equation}
\frac{\Delta\omega}{\omega_o}\simeq-\frac{\overline{\Delta\delta}}{\omega_o^2\int_{r'_1}^{R^*}c^{-2}\kappa_o^{-1}dr'}
\label{eq:scfreqshift}\end{equation}
where $\overline{\Delta\delta}$ is the integral phase shift
\begin{equation}
\overline{\Delta\delta}=\frac{\int_0^{2\pi}\int_0^\pi\Delta\delta(Y^m_l)^2\sin\theta d\theta d\phi} {\int_{0}^{2\pi}\int_0^\pi(Y_l^m)^2\sin\theta d\theta d\phi}.
\label{eq:phaint}\end{equation}
In Equation (\ref{eq:scfreqshift}) $R^*$ is the distance from the stellar center to the base of the starspot, ($r'$, $\theta$, $\phi$) are the spherical coordinates, $Y_l^m$ is a spherical harmonic of degree $l$ and order $m$, $r^{'}_{1}$ is the lower turning point of the mode and the index \textit{'o'} indicates the equilibrium quantities.

The modes that we are considering have $l=0$ and $m=0$, hence, $Y_0^0=1/\left(2\sqrt{\pi}\right)$ and the normalization factor is 
$$\int_0^{2\pi}\int_0^{\pi}\left(Y_l^m\right)^2\sin\theta d\theta d\phi=1.$$

The phase difference $\Delta\delta$ is different from zero only inside the sunspots. The total effect on the frequencies can thus be computed as the superposition of the contributions from each sunspot. Since it is impossible to model each sunspot separately we will instead adopt the $\Delta\delta$ computed for the sunspot considered in this work as a proxy, assuming all sunspots to be alike. 

For a given sunspot, if we centre the coordinate system on the spot's axis, the phase difference is independent of $\phi$ and we find
\begin{equation}
\dfrac{\Delta \omega}{\omega_o}\simeq-\dfrac{1}{2\omega_o^2}\dfrac{\int_o^{\theta_{max}}\Delta\delta(\theta)\sin\theta d\theta}{\int_{r'_1}^{R^*}c^{-2}\kappa_o^{-1}dr'},
\end{equation}
where $\theta_{max}$ coincides with the edge of the spot. 

The integral in the denominator extends only from the centre to the base of the sunspot. In that region spherical symmetry is maintained and the integral is best computed in spherical coordinates. To compute the integral in the numerator, we convert to cylindrical coordinates. Noting that the ratio between $r$ and $R^*$ is equal to the sine of $\theta$, which is a very small angle, we write $\theta\simeq \dfrac{r}{R^*}=\tilde{r}$ and find, for a given sunspot,
\begin{equation}
\dfrac{\Delta\omega}{\omega_o}\simeq-\dfrac{1}{2\omega_o^2}\dfrac{\int_{0}^{\tilde{r}_{max}} \Delta\delta\tilde{r}d\tilde{r}}{\int_{r'_1}^{R^*}c^{-2}\kappa_o^{-1}dr'},
\end{equation}
where $\tilde{r}_{max}$ is the ratio between $r$ at the spot's outer boundary of the penumbra and $R^*$.

Assuming that all spots are alike, and keeping in mind that we are considering modes of degree $l=0$, for which the weighing by the spherical harmonic is independent of the spot's position, we find that the contribution to the frequency shifts of a number $N$ of sunspots combined is 
\begin{equation}
\dfrac{\Delta\omega}{\omega_o}\simeq-\dfrac{N}{2\omega_o^2}\dfrac{\int_{0}^{\tilde{r}_{max}}\Delta\delta\tilde{r} d\tilde{r}}{\int_{r'_1}^{R^*}c^{-2}\kappa_o^{-1}dr}.
\label{eq:freqshift}
\end{equation}

For the case of sun, we estimated $N$ as the ratio between the daily sunspots' area (yearly averaged) at solar maximum and the area covered by our spot. Since the spot used as a proxy in our work is large (40 Mm from the axis to the edge), $N$ estimated in this way is only 3. 

\begin{figure}
\includegraphics[width=8.2cm]{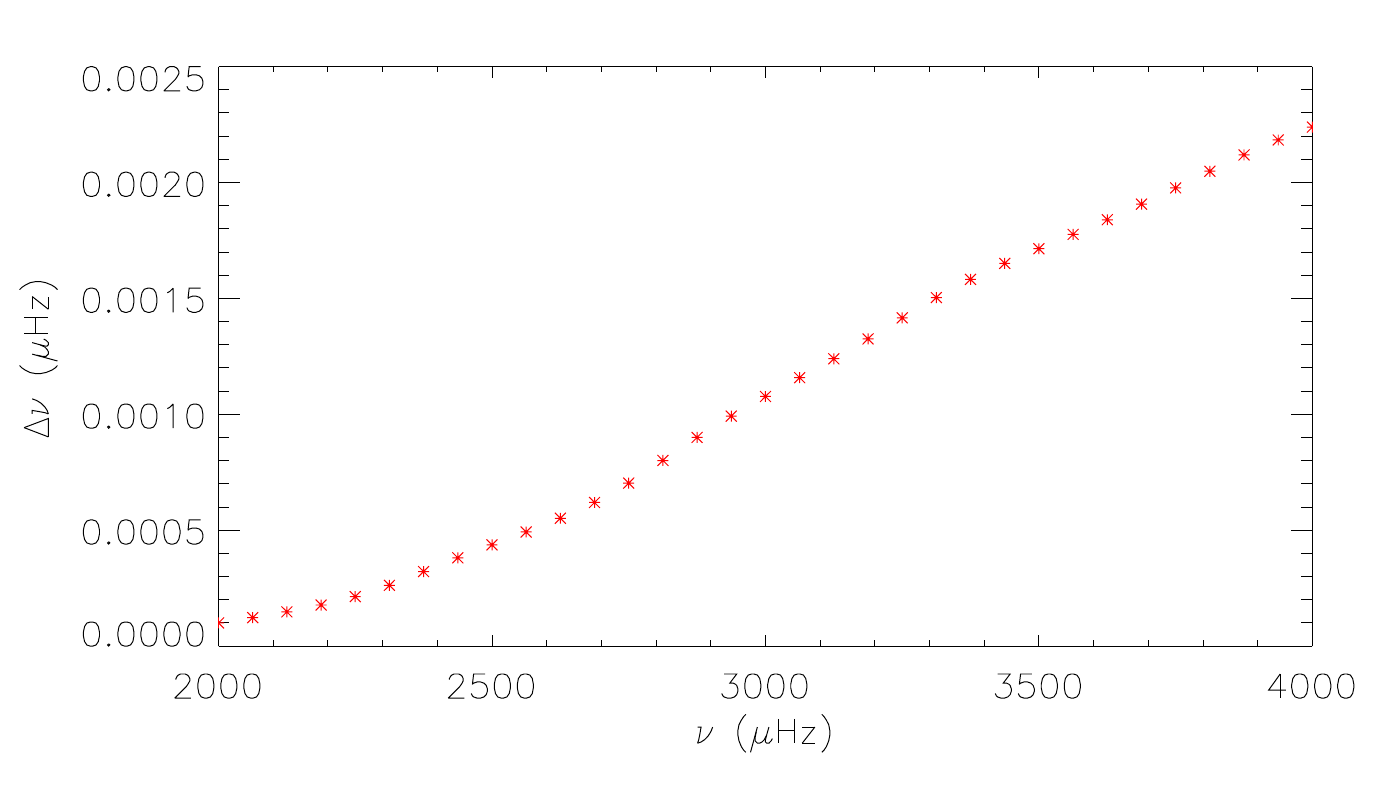}
\caption{Frequency shifts associated with the indirect effect of sunspots considered in this work, as a function of frequency.}
\label{fig:freqshift}
\end{figure}

For $\nu=3711\mu Hz$ we found, using Equation (8), a frequency shift of $~2\times 10^{-3}\mu Hz$. In Figure \ref{fig:freqshift} we show the frequency shifts computed for a range of frequencies.

\section{Conclusions}
\label{sec:conclusions}

We have computed the frequency shifts that result from comparing a spherically symmetric model of the sun and an otherwise similar model but with sunspots. We did not take into account the direct effect of the magnetic field on the oscillations, nor its effect on the local dynamics. The only effect considered in the present work is that resulting from the changes in the stratification within the spots. 

The values of the frequency shifts found are almost two orders of magnitude smaller than the observed frequency shifts between solar maximum and minimum. This indicates that the stratification changes within sunspots do not, by themselves, contribute significantly to the overall frequency shifts along the solar cycle. Since the main contribution of this indirect effect comes from the umbra, the result might increase slightly if we consider the more realistic case of a greater number of smaller spots. Nevertheless, from the present calculation we may already expect this indirect magnetic field contribution to be always significantly smaller that the observed frequency shifts.

Our future objective is to obtain the magnetic solutions and compute the induced frequency shifts in order to estimate the direct effect of starspots to the observed frequencies. From the experience gained in this work we expect that the contribution associated with the direct effect of the magnetic field will be comparable to the observed frequency shifts only if the phase shift associated with the direct effect is of order 1 and extends beyond the umbra. 
\vspace{0.3cm}
\acknowledgements
The authors wish to thank FCT/MCTES, Portugal, for financial support through the project PTDC/CTE-AST/98754/2008. MC is partially funded by POPH/FSE (EC). We are also very grateful to European Science Foundation (ESF) for the travel support and the grant for the conference fee.

\end{document}